\documentclass[prd,aps,twocolumn]{revtex4-2}
\usepackage{epsfig}
\usepackage{graphicx}
\usepackage{color}
\usepackage{hyperref}
\usepackage{ragged2e}

\begin{document}

\title{Energy-Dependent Extension of the Crab Nebula as a Diagnostic of Ultra-High-Energy Hadronic Emission}

\author{Fang-Wu Lu$^1$}
\author{Bo-Tao Zhu$^2$}\email[]{bo\_taozhu@126.com}
\author{Ji-Yang Ren$^1$}
\author{Li Zhang$^3$}\email[]{lizhang@ynu.edu.cn}
\affiliation{$^1$Department of Physics, Yuxi Normal University, Yuxi, 653100, China\\
	$^2$College of Science, Yunnan Agricultural University, Kunming, 650201, China\\
	$^3$Department of Astronomy, Key Laboratory of Astroparticle Physics of Yunnan Province, Yunnan University, Kunming 650091, China}

\date{\today}

\begin{abstract}
The energy-dependent apparent extension of the Crab Nebula, inconsistent with leptonic expectations between X-ray and TeV bands, can be naturally accounted for in a spatially resolved lepto-hadronic scenario. A transition to hadronic dominance occurs above $\sim400~\rm{TeV}$, with nearly pure hadronic emission at PeV energies. The predicted $68\%$ containment radius at $1.1~\rm{PeV}$ is $\sim1.4'$, about six times larger than in purely leptonic scenarios. This energy-dependent extension therefore provides an observable discriminator of the origin 
of ultra-high-energy gamma-ray emission in pulsar wind nebulae, offering a diagnostic of PeV hadronic acceleration in the Crab Nebula.
\end{abstract}


\maketitle

{\it Introduction--}
The Crab Nebula, powered by the Pulsar PSR B0531+21, is the brightest pulsar wind nebula (PWN) in the Galaxy and has been observed across the entire electromagnetic spectrum, from MHz radio waves to ultra‑high‑energy (UHE; $E_{\gamma}\gtrsim100~{\rm TeV}$) gamma‑rays. Synchrotron emission extending up to $\sim100~\rm{MeV}$, detected by the Fermi-LAT and AGILE \cite{Abdo2011Sci,Tavani2011Sci}, together with gamma-ray photons reaching $1.1~\rm{PeV}$ observed by the Large High Altitude Air Shower Observatory (LHAASO) \cite{LhaasoCollaboration2021Sci}, indicate particle acceleration up to PeV energies. These observations establish the Crab Nebula as a natural laboratory for probing particle acceleration and radiation processes in PWNe.

It is widely accepted that relativistic particles in the Crab Nebula are injected at the termination shock (TS) surrounding the central pulsar \cite{Rees1974MNRAS,Kennel1984aApJ}. Emission from radio to X-ray energies is attributed to synchrotron radiation from leptons in the magnetic field, while the GeV–TeV gamma-ray component arises from inverse Compton (IC) scattering of relativistic electrons off ambient photon fields  \cite{Kennel1984bApJ,deJager1992ApJ,Atoyan1996MNRAS,Zhang2008ApJ}, forming the standard leptonic model. 
The recently detected PeV emission, however, challenges this framework, and additional leptonic components beyond the standard scenario have been proposed \cite{LhaasoCollaboration2021Sci}.

Although leptonic models can successfully account for the observed spectral energy distribution (SED) of the Crab Nebula, a non-negligible hadronic contribution to its UHE gamma-ray emission cannot be ruled out. Relativistic protons injected into the pulsar wind may produce TeV–PeV gamma rays through proton–proton interactions and subsequent neutral pion ($\pi^0$) decay \cite{Atoyan1996MNRAS,Yang2009AA,Zhang2020MNRAS,Liu2021ApJ,Nie2022ApJ}. This hadronic channel thus provides an 
additional component alongside the IC emission, leading to a lepto-hadronic hybrid scenario.

Distinguishing between leptonic and hadronic origins of the UHE gamma-ray emission based solely on the multiwavelength spectrum remains challenging. Although neutrino emission would provide a definitive diagnostic of hadronic processes, no significant neutrino signal has yet been detected from the Crab Nebula \cite{Aartsen2017ApJ,Aartsen2020PhRvL,Aartsen2020ApJ,Abbasi2025ApJ}. The predicted neutrino flux from the Crab Nebula is likely below the sensitivity of current instruments, 
such as IceCube and KM3NeT \cite{IcecubeCollaboration2023Sci,Adrian-Martinez2016JPhG}, 
as indicated in Refs.~\cite{Peng2022ApJ,Spencer2025AA}.

Recent studies (see Refs.~\cite{Dirson2023AA,Aharonian2024AA}) have shown that the spatial extension of the Crab Nebula is strongly energy-dependent, exhibiting radial shrinkage in the X-ray and TeV bands. This behavior provides valuable constraints on particle transport and radiative mechanisms.
Within pure leptonic models, it is challenging to simultaneously reproduce the SED and the energy-dependent extension, as demonstrated in Refs.~\cite{Aharonian2024AA,Lu2026ApJ}. This discrepancy is most pronounced in the TeV band, where the measured extension exceeds that predicted by leptonic models, implying that an additional gamma-ray component--such as a hadronic contribution--may be necessary.

In this work, we show that a spatially resolved lepto-hadronic model can simultaneously reproduce the multiwavelength spectrum and apparent extension of the Crab Nebula, demonstrating that the energy-dependent extension provides a direct discriminator of the origin of UHE gamma-ray emission in PWNe.

{\it Spatially resolved lepto-hadronic model--}
A PWN is powered by the spin-down luminosity of its central pulsar \cite{Goldreich1969ApJ,Kennel1984aApJ}. In our model, the total power injected into the PWN from the associated pulsar is distributed among four components:
\begin{equation}
	L(t)=\eta_{\rm B}L(t)+\eta_{\rm e}L(t)+\eta_{\rm p}L(t)+\eta_{\rm T}L(t)\;,
	\label{En_d}
\end{equation}
where $L(t)$ is the spin-down luminosity, and $\eta_{j}$ ($j$= B, e, p, T) represent the energy fractions converted into magnetic field, electron plasma, proton plasma, and turbulent waves, respectively, with $\eta_{\rm B}+\eta_{\rm e}+\eta_{\rm p}+\eta_{\rm T}=1$. The temporal evolution of $L(t)$ is given by \cite{Pacini1973ApJ}
\begin{equation}
	L(t)=L_{\rm sd}
	\left(\frac{1+t/\tau_0}
	{1+T_{\rm age}/\tau_0}\right)^{-\frac{n+1}{n-1}}\;,
	\label{Lsd}
\end{equation}
where $n$ is the braking index, $L_{\rm sd}$ is the current luminosity, $T_{\rm age}$ is the pulsar age, and $\tau_0=2\tau_{\rm c}/(n-1)-T_{\rm age}$ is the initial spin-down timescale, with $\tau_{\rm c}$ being the characteristic age of the pulsar.

The evolution of particles in a spherically (or quasi-spherically) 
symmetric PWN is governed by the advection–diffusion equation
\begin{eqnarray}
	\nonumber
	\frac{\partial n_i}{\partial t}
	&=&\frac{1}{r^2}\frac{\partial}{\partial r}
	\left[r^2 D_i \frac{\partial n_i}{\partial r}\right]
	-\frac{1}{r^2}\frac{\partial}{\partial r}
	\left[r^2 v_{\rm f} n_i\right] \\
	&&-\frac{\partial}{\partial \gamma_i}
	\left[\dot{\gamma}_i n_i\right]
	+ Q_i \delta(r-r_{\rm ts}),
	\label{nep}
\end{eqnarray}
where $n_i=n_i(r,\gamma_i,t)$ is the differential particle number density for species $i$, with $i=e,p$ denoting electrons and protons, $r$ represents the radial position, $\gamma_i$ describes the particle Lorentz factor.

The diffusion coefficient is calculated by $D_i=r_{i,{\rm g}}c\,u_{\rm B}/[3k_{i,{\rm res}}w_{\rm T}(k_{i,{\rm res}})]$, within the quasi-linear approximation \cite{Skilling1975MNRAS}, where 
$u_{\rm B}=B^2(r,t)/8\pi$ is the magnetic field energy density, 
$r_{i,{\rm g}}$ is the particle gyroradius, and $w_{\rm T}(k_{i,{\rm res}})$ is the turbulent wave spectrum evaluated at the resonance wavenumber of $k_{i,{\rm res}}$ \cite{Ptuskin2003AA,Evoli2018PhRvL}. The magnetic field is modeled as 
$B(r,t)=B_{0}(t)[r/r_{\rm ts}(t)]^{\alpha_{\rm mag}}$, where $\alpha_{\rm mag}$ is the magnetic field gradient index, $r_{\rm ts}(t)$ is the TS radius, and $B_0(t)$ is the magnetic field strength at the shock, which is obtained from magnetic-energy conservation in the nebula \cite{Pacini1973ApJ}. Further details 
are given in Ref.~\cite{Lu2026ApJ}.

The advection velocity is parameterized as 
$v_{\rm f}=v_{\rm pwn}(t)\,[r/r_{\rm pwn}(t)]^{\alpha_{\rm adv}}$, 
assuming that the flow velocity matches the dynamical velocity of the nebula at the outer boundary, i.e., $v_{\rm f}(r_{\rm pwn})=v_{\rm pwn}(t)$. 
Here, $v_{\rm pwn}(t)$ and $r_{\rm pwn}(t)$ are the expansionl velocity and radius of the nebula at time $t$, respectively, and $\alpha_{\rm adv}$ denotes the velocity gradient index. 
In addition, the energy-loss term $\dot{\gamma}_{i}$ includes adiabatic losses for both particle species, synchrotron and IC losses for electrons, and proton–proton interaction losses for protons.

The injection term $Q_i$ represents particle injection at the TS. 
For electrons, we adopt a broken power-law spectrum motivated by the presence of distinct low- and high-energy components in the multiwavelength nonthermal emission of the Crab Nebula \cite{Atoyan1996MNRAS,Zhang2008ApJ}:
\begin{equation}
	Q_{\rm e}=Q_{\rm e,0}(t)
	\left\{
	\begin{array}{ll}
		\left(\frac{\gamma_{\rm e}}{\gamma_{\rm e,b}}\right)^{-\alpha_{\rm e,1}} & \gamma_{\rm e}<\gamma_{\rm e,b}\\
		\left(\frac{\gamma_{\rm e}}{\gamma_{\rm e,b}}\right)^{-\alpha_{\rm e,2}} & \gamma_{\rm e}\geq\gamma_{\rm e,b}
	\end{array}
	\right.,
\end{equation}
where $\gamma_{\rm e,b}$ is the break Lorentz factor separating the low- and high-energy components, and $\alpha_{\rm e,1}$ and $\alpha_{\rm e,2}$ are the corresponding spectral indices below and above the break, respectively. 
This form reflects the distinct particle populations responsible for the radio–GeV and X-ray–TeV emissions.

For protons, we adopt a relativistic Maxwellian distribution with a power-law tail, as suggested by particle-in-cell (PIC) simulations \cite{Spitkovsky2008ApJ,Sironi2011ApJa,Sironi2011ApJb}. 
This form reflects the thermalized particle population downstream of the shock together with a nonthermal component accelerated to high energies:
\begin{equation}
	Q_{\rm p}=Q_{\rm p,0}(t)
	\left\{
	\begin{array}{ll}
		\gamma_{\rm p}\exp\left(-\frac{\gamma_{\rm p}}{\Delta\gamma_{\rm p}}\right) & {\rm for}~\gamma_{\rm p}<\gamma_{\rm p,b}\\
		\gamma_{\rm p}\exp\left(-\frac{\gamma_{\rm p}}{\Delta\gamma_{\rm p}}\right)+\xi\gamma_{\rm p}^{-\alpha_{\rm p}} & {\rm for}~\gamma_{\rm p}\geq\gamma_{\rm p,b}
	\end{array}
	\right. ,
	\label{Qpinj}
\end{equation}
where $\Delta\gamma_{\rm p}$ characterizes the width of the relativistic Maxwellian, $\gamma_{\rm p,b}=$(3-7)$\Delta\gamma_{\rm p}$ denotes the energy at which the power-law component begins, $\alpha_{\rm p}\sim2.4$ represents the spectral index, and $\xi$ is determined from the continuity condition with $\gamma_{\rm p,b}\exp(-\gamma_{\rm p,b}/\Delta\gamma_{\rm p})=\xi\gamma_{\rm p,b}^{-\alpha_{\rm p}}$. Moreover, the normalization factor $Q_{i,0}(t)$ for both particle species is determined from the pulsar energy budget through
$\eta_i L(t)=\int_0^{\infty}\gamma_i m_i c^2 Q_i(\gamma_i,t)\,d\gamma_i$. 

The evolution of the turbulent waves within the nebula is given by~\cite{Lagage1983AA,Evoli2018PhRvL}
\begin{eqnarray}
	\nonumber
	\frac{\partial{w_{\rm{T}}}}{\partial{t}}
	&=&\frac{\partial}{\partial{k}}\left[k^2D_{\rm{kk}}\frac{\partial}{\partial{k}}\left(\frac{w_{\rm{T}}}{k^2}\right)\right]+\Gamma_{\rm{cr}}w_{\rm{T}}-\frac{\partial}{\partial{k}}\left[\dot{k}_{\rm ad}w_{\rm{T}}\right]\\
	&&-\frac{1}{r^2}\frac{\partial}{\partial{r}}\left[r^2v_{\rm{T}}{w_{\rm{T}}}\right]+Q_{\rm{T}}\delta(r-r_{\rm{ts}})\;,
	\label{wk}
\end{eqnarray}
where $w_{\rm{T}}=w_{\rm{T}}(r,k,t)$ is the spectral energy density of turbulent waves, $k$ is the wavenumber,  $\Gamma_{\rm{cr}}$ is the growth rate of the cosmic‑ray streaming instability \cite{Lagage1983AA,Evoli2018PhRvL},  $\dot{k}_{\rm ad}$ represents the adiabatic shift of the wavenumber caused by nebular expansion, and $v_{\rm{T}}{\equiv}v_{\rm{f}}$ denotes the advection velocity of turbulence, assumed to follow the background flow. The diffusion coefficient in the  Kolmogorov-type cascade is taken as $D_{\rm{kk}}=k^{3}\nu_{\rm{A}}[\frac{1}{2}kw_{\rm{T}}/u_{\rm{B}}]^{1/2}$, with $\nu_{\rm A}$ denoting the local Alfv\'{e}n velocity \cite{Zhou1990JGR,Miller1996ApJ}. $Q_{\rm T}=Q_{\rm T,0}(t)\delta(k-k_{\rm inj})$ describes monochromatic injection at the TS with a wavenumber of $k_{\rm inj}=1/r_{\rm ts}(t)$, where $Q_{\rm T,0}(t)$ is determined from $\eta_{\rm T}L(t)=\int_{0}^{\infty}Q_{\rm{T}}dk$.

The coupled equations for particles and turbulent waves are solved 
numerically using a finite-difference Crank–Nicolson method. Nonthermal photon emission from synchrotron radiation, IC scattering off ambient photon fields, and proton–proton interactions is calculated following 
Refs.~\cite{Blumenthal1970RvMP,Atoyan1996MNRAS,Kelner2006PhRvD}. For thermal dust emission, we adopt the same treatment as in Ref.~\cite{Dirson2023AA}. The spatial profile of the emission is calculated using the method described in Ref.~\cite{Aharonian2024AA}.

\begin{figure*}
	\centering
	\includegraphics[width=0.96\textwidth]{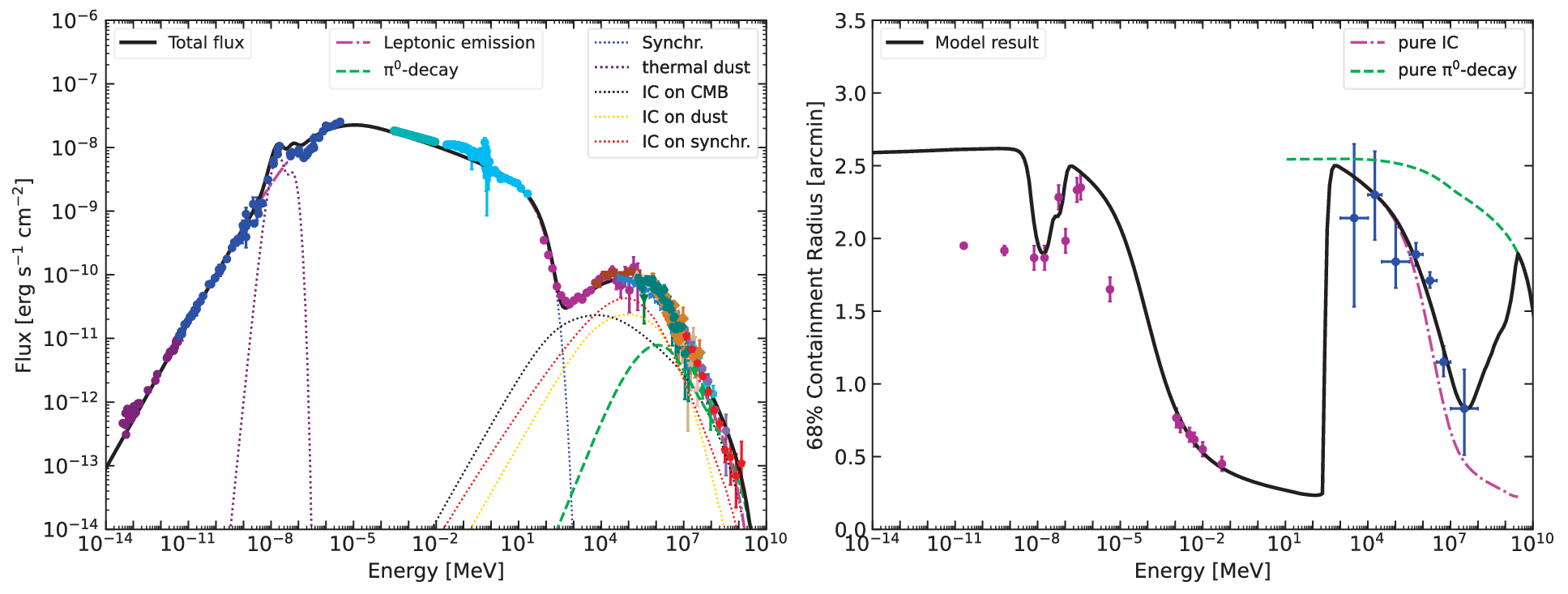}
	\caption{Left panel: Multiwavelength spectrum of the Crab Nebula. The black solid line shows the total emission, while the magenta dash-dotted and green dashed lines represent the leptonic and hadronic ($\pi^{0}$-decay) components, respectively. The dotted lines indicate the contributions from synchrotron radiation (blue), thermal dust emission (purple), and inverse Compton (IC) scattering off the cosmic microwave background (CMB; black), dust (yellow), and synchrotron photons (red). 
		Right panel: Modeled $68\%$ flux containment radius (black solid line). The magenta dash-dotted line shows the gamma-ray extension predicted in the pure leptonic (IC) scenario, while the green dashed line corresponds to the hadronic $\pi^{0}$-decay scenario.}
	\label{Fig1}
\end{figure*}

\begin{table}
	\centering
	\caption{Model parameters adopted and fitted to reproduce the spectrum and spatial extension of the Crab Nebula.}
	\label{tab:1}
	\setlength{\tabcolsep}{1.1mm}
	\begin{ruledtabular}
		\begin{tabular}{lccc}
			Parameter & Symbol & Adopted & Ref.~\cite{Lu2026ApJ}\\
			\colrule
			\multicolumn{4}{l}{\textit{Leptonic injection}$^a$} \\
			Electron efficiency & $\eta_{\rm e}$ & 0.28 & 0.26\\
			Magnetic efficiency & $\eta_{\rm B}$ & 0.02 & 0.02\\
			Electron low-energy index & $\alpha_{\rm e,1}$ & 1.45 &1.42\\
			Electron high-energy index & $\alpha_{\rm e,2}$ & 2.23 &2.25\\
			Electron break Lorentz factor  & $\gamma_{\rm e,b}$ & $8.9\times10^5$ &$8.9\times10^5$\\
			Electron max Lorentz factor & $\gamma_{\rm e,max}$ & $5.6\times10^9$ &$5.6\times10^9$\\
			\colrule
			\multicolumn{4}{l}{\textit{Radial indices}$^a$} \\
			Advection index & $\alpha_{\rm adv}$ & $-1.10$ &-1.19\\
			Magnetic-field index & $\alpha_{\rm mag}$ & $-0.28$ &-0.30\\
			\colrule
			\multicolumn{4}{l}{\textit{Hadronic injection}$^b$} \\
			Proton efficiency & $\eta_{\rm p}$ & 0.25 &$\cdots$\\
			Proton Maxwellian width & $\Delta\gamma_{\rm p}$ & $9.0\times10^3$ &$\cdots$\\
			Proton break Lorentz factor & $\gamma_{\rm p,b}$ & $4\Delta\gamma_{\rm p}$&$\cdots$\\
			Proton max Lorentz factor & $\gamma_{\rm p,max}$ & $1.6\times10^7$ &$\cdots$\\
		\end{tabular}
	\end{ruledtabular}
	\begin{minipage}{\linewidth}
		\RaggedRight
		\footnotesize
		$^a$ The adopted parameters are slightly adjusted from Ref.~\cite{Lu2026ApJ} to consistently reproduce the multiwavelength emission from radio to TeV energies and the observed spatial extent by incorporating the hadronic injection. \\
		$^b$ Determined by fitting the observed TeV--PeV gamma-ray spectrum and nebular extension.
	\end{minipage}
\end{table}

{\it Results--} To investigate the hadronic contribution to the UHE gamma-ray emission from the Crab Nebula, we simultaneously reproduce its multiwavelength spectrum and energy-dependent spatial extension ($68\%$ flux containment radius). We adopt the multiwavelength observational data compiled by Ref.~\cite{Lu2026ApJ} and the measurements of the spatial extension reported by Ref.~\cite{Aharonian2024AA}. 
The dynamical evolution of the nebula is modeled following Ref.~\cite{Bandiera2023MNRAS}, yielding the present-day TS and nebular radii consistent with observations \cite{Weisskopf2012ApJ}. The target proton density is estimated 
from the nebular mass and volume, yielding a characteristic value of $n_{\rm tar}\sim10~\mathrm{cm^{-3}}$ \cite{Zhang2020MNRAS,LhaasoCollaboration2021Sci}.

The model parameters used in this work are summarized in Table~\ref{tab:1}. The leptonic parameters are taken from 
Ref.~\cite{Lu2026ApJ}, while the hadronic parameters are determined by fitting the TeV--PeV gamma-ray spectrum and spatial extension. The radial indices of the magnetic field and 
advection velocity are slightly adjusted from those in Ref.~\cite{Lu2026ApJ} to reproduce the observed energy-dependent extension in the presence of hadronic components. 

Fig.~\ref{Fig1} shows the modeled multiwavelength spectrum and spatial extension of the Crab Nebula. To highlight the difference in the $68\%$ flux containment radius between leptonic and hadronic gamma-ray production, the right panel presents the results calculated separately in pure leptonic and pure hadronic frameworks. The predicted radius in the pure hadronic case is larger than the observed values. In contrast, the pure leptonic case falls below the observed data at TeV energies---a result consistent with the findings of Refs.~\cite{Aharonian2024AA,Lu2026ApJ}. 

This behavior arises because higher-energy electrons experience more severe energy-dependent synchrotron losses during propagation inside the nebula \cite{Lu2026ApJ}, whereas the hadronic cooling time in proton--proton interactions is nearly energy-independent above $\sim1~{\rm GeV}$ \cite{Aharonian2004vhecbook}, so that the high-energy proton distribution is mainly governed by transport. As a result, the electron population decreases more rapidly with distance from the TS than the proton population, leading to a steeper electron spectrum at high energies. This effect ultimately results in a more compact 
high-energy photon emission region in purely leptonic scenarios. Consequently, a lepto-hadronic hybrid model provides a more natural explanation for the TeV--PeV gamma-ray emission from the Crab Nebula. In particular, it helps reconcile the discrepancy 
between the observed TeV spatial extension and that predicted by pure leptonic models, and naturally accounts for the difference between the TeV and X-ray extensions.

\begin{figure}
	\centering
	\includegraphics[width=0.47\textwidth]{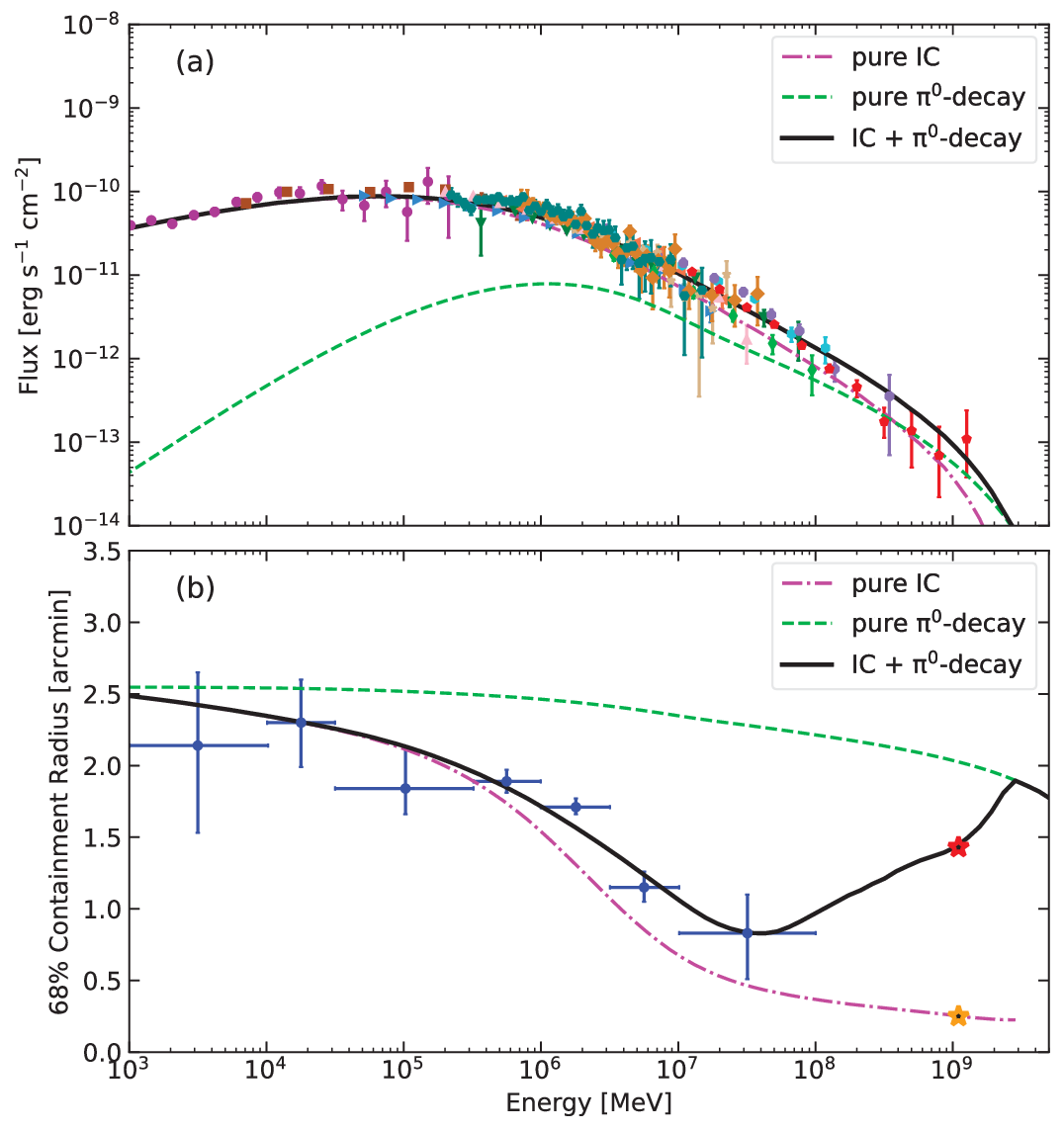}
	\caption{Upper panel: Gamma-ray spectrum of the Crab Nebula. 
	The total spectrum (black solid line) is decomposed into the IC component from leptons (magenta dash-dotted line) and the hadronic $\pi^0$-decay component (green dashed line). 
	Lower panel: The $68\%$ flux containment radius as a function of gamma-ray energy for three cases: a pure leptonic model, a pure hadronic model, and a lepto-hadronic hybrid model. 
	The orange star marks the predicted spatial extension radius at $1.1~{\rm PeV}$ for the pure leptonic model, while the red star corresponds to the lepto-hadronic hybrid model.}
	\label{Fig2}
\end{figure}

\begin{figure}
	\centering
	\includegraphics[width=0.47\textwidth]{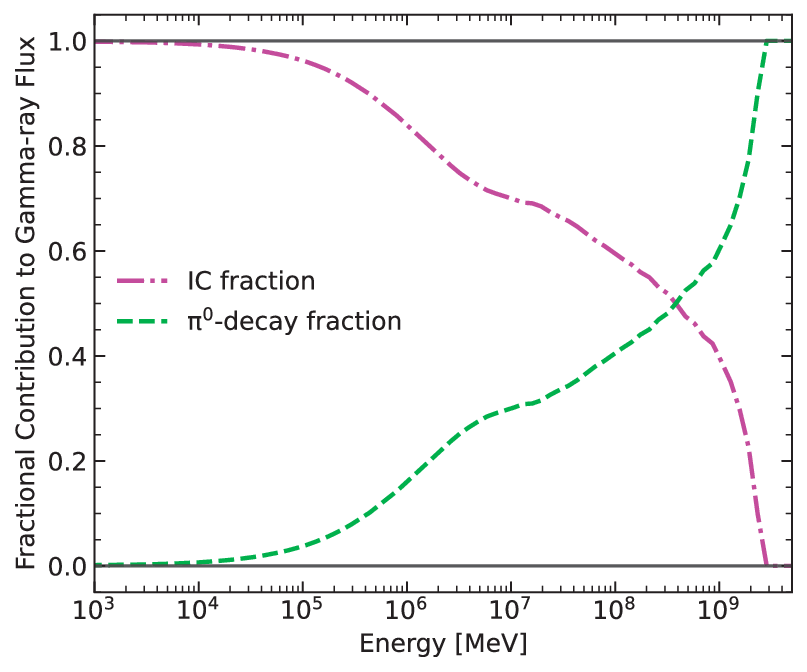}
	\caption{Flux contribution fractions of the leptonic (IC; dash-dotted line) and hadronic ($\pi^0$-decay; dashed line) components to the gamma-ray emission of the Crab Nebula.}
	\label{Fig3}
\end{figure}

Fig.~\ref{Fig2} provides a zoomed view of the gamma-ray spectrum and spatial extension profile. The fractional contributions of leptonic and hadronic components to the gamma-ray flux are shown in Fig.~\ref{Fig3}. These results highlight a clear 
transition from leptonic-dominated emission at GeV energies to a substantial hadronic contribution at TeV energies. At TeV energies—particularly above $\sim10~{\rm TeV}$—hadrons contribute an increasing fraction of the flux and significantly affect the spatial extension. We find that the hadronic $\pi^0$-decay contribution rises from $\sim15\%$ at $1~{\rm TeV}$ to $\sim50\%$ at $\sim400~{\rm TeV}$, and approaches $100\%$ at PeV energies.

The PeV gamma-ray emission from the Crab Nebula detected by LHAASO is naturally reproduced by our lepto-hadronic hybrid model. In this model, the emission at PeV energies is dominated by hadronic $\pi^0$-decay, consistent with the suggestions of Refs.~\cite{Liu2021ApJ,Peng2022ApJ,Nie2022ApJ}. In our model, the predicted $68\%$ flux containment radius at $1.1~{\rm PeV}$ is about $1.43'$, significantly larger than the 
$\sim0.25'$ expected in a pure leptonic scenario. This pronounced difference in spatial extension suggests that the energy-dependent extension can serve as an effective diagnostic for distinguishing the origin of PeV photon emission in the Crab Nebula—a prediction that will be testable with the upcoming Large Array of Imaging Atmospheric Cherenkov Telescopes (LACT; see Ref.~\cite{Zhang2024icrc}).

{\it Conclusions--}
We study the TeV--PeV gamma-ray emission from the Crab Nebula within a lepto-hadronic hybrid model. As shown in Figs.~\ref{Fig1} and \ref{Fig2}, the inclusion of hadronic components helps reconcile the apparent discrepancy between the measured extensions in the TeV and X-ray bands. In our model, protons provide a substantial contribution to the gamma-ray flux at 
$\gtrsim10~{\rm TeV}$, which becomes increasingly hadron-dominated as the energy rises to about $400~{\rm TeV}$ (see Fig.~\ref{Fig3}). This result lends support to the suggestion by Ref.~\cite{Atoyan1996MNRAS} that hadronic nuclear collisions can make a significant contribution to the gamma-ray spectrum at energies above $\sim10~{\rm TeV}$.

Our model allocates about $25\%$ of the spin-down luminosity to proton components, consistent with the estimates of Refs.~\cite{Liu2021ApJ,Peng2022ApJ}. 
However, while those studies associate the proton contribution primarily with PeV-range emission, our results indicate that a substantial hadronic component is already present at TeV energies. The PeV photon emission from the Crab Nebula 
in our model is likewise dominated by hadronic $\pi^0$-decay.

Finally, our model predicts that if the $1.1~{\rm PeV}$ photons originate from hadrons injected at the TS, the corresponding $68\%$ flux containment radius would be about $1.4'$, approximately six times larger than expected in a pure leptonic scenario. This pronounced difference in spatial extension provides a clear observational diagnostic for distinguishing the leptonic and hadronic origins of UHE gamma-ray emission in PWNe.

\begin{acknowledgments}
We thank the anonymous referee for helpful comments. F.W.L. is supported by the National Natural Science Foundation of China (NSFC) under Grant No. 12363006. 
L.Z. is supported by NSFC under Grant No. 12233006. 
B.T.Z. acknowledges support from NSFC under Grant No. 12363007; the Yunnan Fundamental Research Projects 
(Grant No. 202501AT070152); and the Xingdian Talent Support Program of Yunnan Province.
\end{acknowledgments}

\bibliography{msref.bib}

\end{document}